# Non-linear relationship of cell hit and transformation probabilities in low dose of inhaled radon progenies

**Imre Balásházy[1,2], Árpád Farkas[1], Balázs Gergely Madas[1] and Werner Hofmann[3]**

[1]*Hungarian Academy of Sciences KFKI Atomic Energy Research Institute*
*Health and Environmental Physics Department*
*1525 Budapest, P.O. Box 49, Hungary*

[2]*Aerohealth Scientific Research Development and Servicing Ltd. Co.*
*2090 Remeteszőlős, Csillag sétány 7, Hungary*

[3]*University of Salzburg, Division of Physics and Biophysics*
*Department of Materials Engineering and Physics*
*5020 Salzburg, Hellbrunner Str. 34, Austria*

**Abstract.** Cellular hit probabilities of alpha particles emitted by inhaled radon progenies in sensitive bronchial epithelial cell nuclei were simulated at low exposure levels to obtain useful data for the rejection or in support of the linear-non-threshold (LNT) hypothesis. In this study, local distributions of deposited inhaled radon progenies in airway bifurcation models were computed at exposure conditions, which are characteristic of homes and uranium mines. Then, maximum local deposition enhancement factors at bronchial airway bifurcations, expressed as the ratio of local to average deposition densities, were determined to characterize the inhomogeneity of deposition and to elucidate their effect on resulting hit probabilities. The results obtained suggest that in the vicinity of the carinal regions of the central airways the probability of multiple hits can be quite high even at low average doses. Assuming a uniform distribution of activity there are practically no multiple hits and the hit probability as a function of dose exhibits a linear shape in the low dose range. The results are quite the opposite in the case of hot spots revealed by realistic deposition calculations, where practically all cells receive multiple hits and the hit probability as a function of dose is non-linear in the average dose range of 10-100 mGy.

# 1. Introduction

The characterization of biological effects of the exposure at low doses of ionizing radiation seems to be one of the most challenging tasks of current radiation protection, radiation biophysics and radiation biology. The most controversial issue of this research field is the question of the linear-non-threshold (LNT) dose-effect hypothesis, which suggests that any infinitesimal radiation dose can be detrimental and hence the linear relationship between radiation dose and related adverse health effects is valid even in the range of low doses. The validity of the application of the LNT relationship is supported by national and international organizations, such as the International Commission on Radiological Protection (ICRP 2006), the National Commission on Radiological Protection (NCRP 2001), the United Nations Scientific Committee on the Effects of Atomic Radiation (UNSCEAR 2006), and the Commission on the Biological Effects of Ionizing Radiation VII (BEIR VII 2005). Thus, current radiological protection recommendations are based on the assumption that the risk of adverse health effects is proportional to radiation exposure down to zero dose. This recommendation has been questioned by many scientists (e.g. Becker 1997, Bond et al. 1996, Cohen 2000, 2008, Jaworowski 1999, Sagan 1989, Stewart and Kneale 1990, Tubiana 2000) and the French Academy of Sciences (Tubiana et al. 2005) and it is one of the most debated scientific issues of the last decades.

Considerable efforts have been spent and a large variety of methods applied to obtain scientific evidence for the rejection or in support of the LNT hypothesis (Charles 2006). Epidemiological studies of atomic bomb survivors (Thompson et al. 1994, Preston et al. 1994, 2007, Pierce et al. 1996, Pierce and Preston 2000) and former uranium miners (BEIR VI 1999, Tomasek et al. 2008, Brueske-Hohlfeld et al. 2006, Lubin et al. 1995, Mulloy et al. 2001) revealed that there was a strong linear correlation between high radiation exposure levels and enhanced lung cancer incidences, leukaemia, thyroid cancer and other health disorders. However, direct extrapolation of atomic bomb and miner data to the dose ranges general population is exposed to may be questionable due to the different exposure conditions. For instance, uranium mines are characterized by equilibrium factors, unattached fractions and aerosol size distributions, significantly differing from those characterizing the homes or the outdoor environment (NRC 1991). Such an extrapolation could also be challenged by the different dose-rates and types of radiation involved, as well as by non-targeted effects, e.g. bystander effects (Hall 2003, Brenner et al. 2001, Morgan 2006, Little M P 2004), which are significant at low and very low doses. At the same time, in spite of the sophisticated analyses, residential radon studies yielded statistically poor evidences at low doses (Alavanja et al. 1994, Kreienbrock et al. 2001, Lubin et al. 2004, Darby et al. 2005, Krewski et al. 2005, Wichmann et al. 2005).

Furthermore, experimental radiation carcinogenesis has not yet provided a plausible dose-effect relationship in the low-dose range (Chadwick and Leenhouts 2002, Duport 2003). This suggests that the understanding of mechanisms of action of radiation and cellular changes which may eventually lead to malignancy is becoming increasingly important in biological risk estimation (Clarke 2001). Hence, the development and application of mechanistic models of radiation carcinogenesis seems to be a promising avenue. However, mechanistic risk models are either based on the study of alpha tracks or on exact microdosimetric quantities, like cellular doses or alpha-hit frequencies (Klebanov et al. 1993, Crawford-Brown and Hofmann 2002, Chadwick et al. 2003). In case of inhalable radioisotopes, such as radon or thoron progenies, these parameters can be obtained through the application of lung deposition and dosimetry models.

Most of the current models simulating the effect of inhaled radionuclides are based on the assumption that the deposited particles are uniformly distributed along the airways (Nikezic et al. 2002, Bohm et al. 2003). However, there is experimental evidence that deposition patterns are highly non-uniform within airway bifurcations (Martin and Jacobi 1972, Kinsara et al 1995, Kim and Fisher 1999). Hence, adverse health effects of inhaled radioactive particles may be related to local burdens rather than to average lung burdens. Indeed, radon progeny induced lung cancer initiations could be localised exactly at the peak of the bifurcations (Schlesinger and Lippmann 1978, Churg and Vedal 1996).

The objectives of the current research are (1) to describe and quantify local radon progeny deposition distributions in the large central airways, (2) to compute related microdosimetric quantities, and (3) to analyse the effect of non-uniform deposition patterns on alpha-hit probabilities, cellular radiation doses and transformation probabilities and resulting potential health consequences. The results obtained may then serve as input parameters for mechanistic models of radiation carcinogenesis. They may also contribute to the elucidation of low dose health effects and hence provide useful information regarding the validity of the LNT hypothesis.

## 2. Methods

The majority of the natural radiation burden is provided by inhaled radon progenies whereby the lung is the most exposed organ in the human body. Consequently, our modelling efforts are focused on the quantification of cellular radiation burden in the lung epithelium following the inhalation of radon progenies. Histological studies conducted by Saccomanno et al. (1996) revealed that radon-induced small-cell lung cancer in humans arises primarily in the segmental and subsegmental airways. This finding is in line with the results of the earlier histological study of Garland (1961), who stated that the occurrence of alpha-particle induced pre-neoplastic and neoplastic lesions is highest in the large bronchi. Later, histopathological investigations revealed that large bronchi were the preferential locations of not only the small-cell lung cancers, but also of the squamous carcinomas (Kreuzer et al. 2000). These findings are in line with our earlier simulation results showing that the deposition densities are the highest in the large bronchi (generations 2-5), exactly where the neoplastic lesions were found (Hofmann et al. 2006). Thus, in the current study, the targeted region of the airways is the large bronchi. Numerical and analytical approaches have been implemented to simulate the transport and airway deposition of attached and unattached radon progenies in large bronchial airways and to characterize the resulting cellular effects of low and intermediate doses of ionizing radiation.

The geometry of the first five generations of the human tracheobronchial airways was created in CAD (computer aided design) environment based on available morphometrical data (Hegedűs et al. 2004). The digitized geometry was discretized by the construction of an inhomogeneous, unstructured numerical mesh (Farkas et al. 2006), and discrete air velocity and pressure values were computed applying the finite volume Navier-Stokes solver of the FLUENT CFD (computational fluid dynamics) program package. Spherical particles were then injected into the computed airflow field and tracked through the whole airway geometry in order to simulate the transport and deposition of inhaled attached and unattached radon progenies. Breathing and radiation exposure parameters characteristic of homes and uranium mines were derived from earlier publications (ICRP 1994, BEIR VI 1999, NRC 1991, UNSCEAR 2000, Haninger 1997, Reineking et al. 1988, Samet et al. 1989, and Cooper et al. 1973). For the quantification of the distribution of deposited particles within the airway bifurcations, the whole surface was scanned with pre-specified surface area elements and local deposition enhancement factors (EF) were computed as the ratio of deposition density in the unit surface area (patch) to the deposition density in the whole system of five airway bifurcations (Balásházy and Hofmann 2000). Since the distribution of EFs is sensitive to the area of the scanning surface element (Balásházy et al. 2003), choosing a biologically plausible patch size is an important element of the model formulation. There is accumulating evidence that the number of cells that respond to an alpha-particle radiation (alpha-track) is greater than the number of cells traversed by that alpha-particle. Although it is not the scope of this article to explore the role of non targeted effects, for the future it is indicated to consider target sizes, which take into account the possible contribution of the cellular responses in "bystander" cells. In this work, the patch size was set to $1.7 \times 10^{-7}$ m$^2$ (0.412 mm x 0.412 mm), equivalent to a bronchial surface area containing about $10^4$ epithelial cells (Mercer et al. 1994).

In the present work, alpha-hit probabilities were computed in the cell nuclei of the epithelium of large bronchial airways for alpha particles emitted (i) from a uniform distribution of deposited radon progenies and, (ii) from deposition hot spots of realistic, inhomogeneous deposition patterns.

In case of uniform activity distributions, average hit probabilities of alpha particles in sensitive bronchial cell nuclei (basal and secretory cells) in the tracheobronchial tree were computed with a stochastic lung deposition model (Hofmann et al. 2002). The user-supplied Monte Carlo code computes both near wall alpha tracks (traversing only tissue) and far wall alpha tracks (traversing air as well as tissue) (Fakir et al. 2006). For cell nuclei located in the hot spot region, that is, at the carinal ridges of the bifurcations, the contribution of far wall alpha-tracks is very small compared to the contribution of the near-wall alpha-tracks. Thus, in case of the hit probability computations in the hot spots, the far wall components may safely be neglected. Since the size of the selected hot spot patch is small enough compared to the curvature of the surface of the bifurcation, the number of far wall and near wall alpha-tracks emitted from the hot spots are about equal. This means that only the half of the alpha tracks of the deposited radon progenies in the hot spot produces hits in the hot spot cell nuclei.

Hit probabilities were computed in 9 μm diameter spherical sites, representing sensitive bronchial cell nuclei located at various depths in the epithelium. Although several microscopic images of the epithelium suggest that cell nuclei are certainly not spherical (see for instance Farmer and Hay, 1991), from dosimetric point of view the consideration of spherical nuclei is a reasonable model assumption. Since both basal and secretory cells have been considered as the primary progenitor cells of the bronchial carcinomas (Ford and Terzaghi-Howe 1992, Johnson 1995), hit calculations refer to both cell types. Their depth-density distributions were adopted from Mercer et al. (1991). In the present work two different averaging procedures were applied: (i) arithmetic average of the hit probabilities computed for the bronchial basal and secretory cell nuclei, and (ii) weighted average of these hit probabilities based on the measured volumetric ratios of these cell nuclei (Mercer et al. 1991, Hofmann et al. 2000). Based on the average alpha-hits per bronchial cell nucleus, multiple hit probabilities were calculated assuming that the number of hits per bronchial cell nucleus follows a Poisson distribution.

To compute the average cell nucleus doses a dose-exposure conversion factor of 4.8 mGy/WLM has been assumed, based on the earlier calculations of Winkler-Heil and Hofmann (2002). This conversion was adopted both for homes and mines, which is equivalent to a K-factor of 1 (see Table B-11 of BEIR VI Report), where K-factor is the dose per unit radon concentration in homes divided by dose per unit radon concentration in mines. The cell nucleus dose was simulated as the computed adsorbed alpha energy in the nucleus divided by the mass of the cell nucleus, where the size and depth distribution of the nucleus were determined by the Mercel et al. (1991) data.

In the U.S. Environmental Protection Agency 2003 Assessment of Risks from Radon in Homes Report (EPA, 2003) some biologically based models are applied to compute the risk from residential radon exposure and it is concluded that the risk can be about two or four times lower than projected by the BEIR VI model, for smokers and never smokers, respectively. These simulations are based on the Moolgavkar and Luebeck (1990) 2-stage model of carcinogenesis. However, these computations did not take into consideration the local inhomogeneity of burden that is the effects of hot spots.

### 3. Results and discussion

Particle deposition patterns were computed for inhaled radon progenies within the central human airways for the home and mine exposure conditions compiled in Table 1. The resulting deposition patterns of inhaled attached and unattached radon progenies in a five-generation bifurcation model are plotted in Figure 1 for both exposure conditions.

**Table 1.** Parameters of the applied home and mine environments.

| | Home | | Uranium mine | |
|---|---|---|---|---|
| Breathing mode | nose breathing | (ICRP 66, 1994) | nose breathing | (ICRP 66, 1994) |
| Flow rate | 18 l/min | (ICRP 66, 1994) | 50 l/min | (ICRP 66, 1994) |
| Particle diameter | Attached: 200 nm  (Haninger, 1997) Unattached: 1.2 nm  (Reineking et al., 1998) | | Attached: 200 nm  (BEIR VI, 1999) Unattached: 1 nm  (BEIR VI, 1999) | |
| Unattached fraction | 6 % | (Haninger, 1997) | 1 % | (Samet et al., 1989) |
| Activity concentration ratios ($^{218}$Po/$^{214}$Pb/$^{214}$Bi) | 0.58/0.44/0.29 (UNSCEAR, 2000) | | 0.60/0.29/0.21 (BEIR VI, 1999) | |

Deposition distributions of attached progenies are highly non-uniform, consistent with the effect of impaction as the primary deposition mechanism for larger particles and the higher flow velocities in the upper airway generations (Farkas et al. 2006). For the unattached fraction, Brownian motion is the dominating deposition mechanism, causing more scattered deposition patterns (Farkas et al. 2008). The apparent differences between the distributions in homes and mines are the result of the differences in the flow rates and the unattached fractions (see Table 1). For comparison, the bottom panel displays the corresponding uniform distribution where the same amount of deposition sites and alpha activities are homogeneously distributed along a computational mesh of a cylinder. The microdosimetric models applying uniform distribution of deposition and activity approximate the airways by cylinders characterized by lengths and diameters (Winkler-Heil and Hofmann 2002).

The quantification of the deposition distribution has been performed by computing deposition enhancement factors (EF) (Farkas and Balásházy 2008), applying the technique described in the Methods section. Figure 2 illustrates the distribution of EFs for homes (upper panel) and mines (bottom panel).

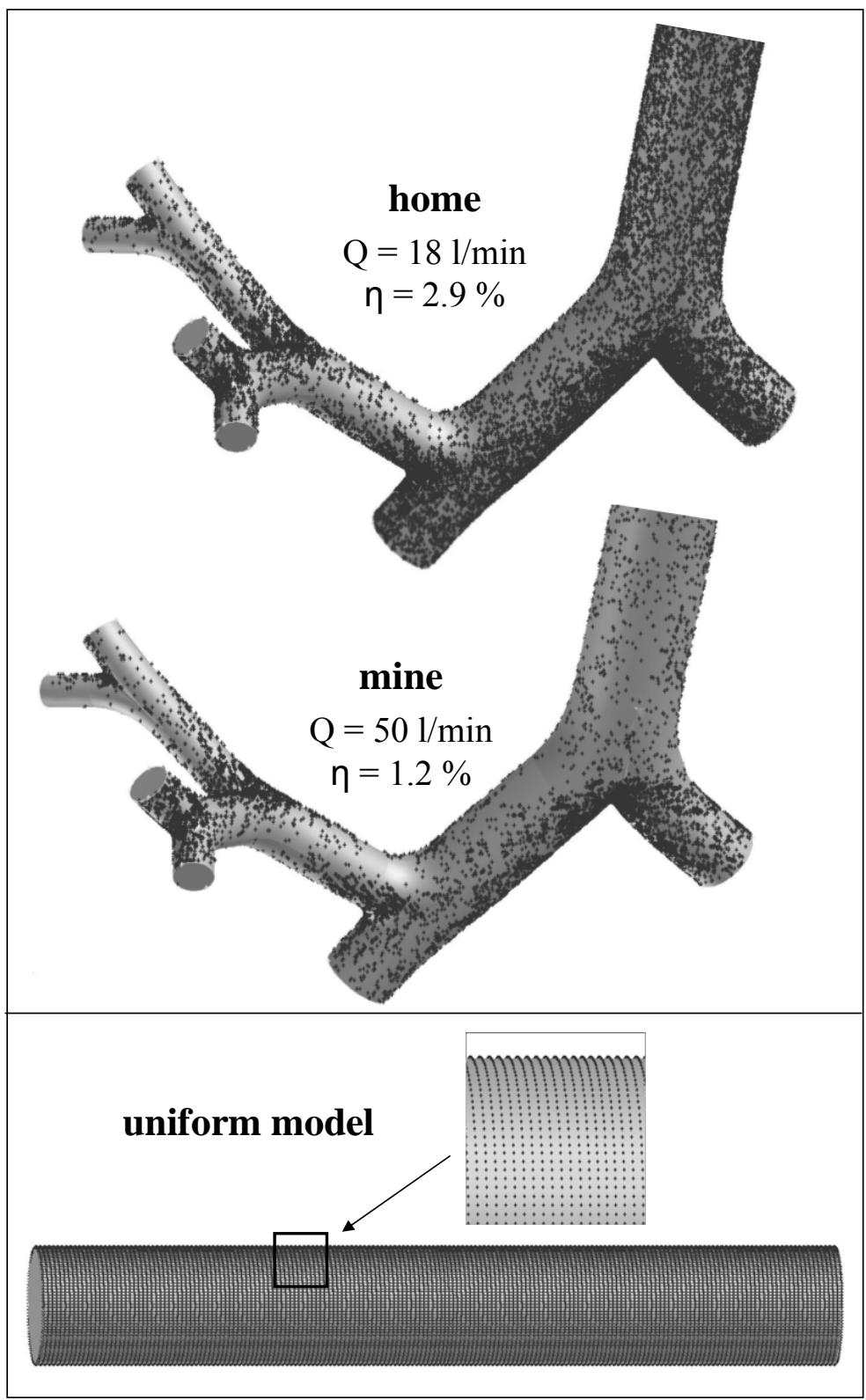

**Figure 1.** Inspiratory deposition patterns of attached and unattached radon progenies under specific home and mine exposure conditions (see Table 1). Q: tracheal flow rate, η: deposition efficiency. In the uniform model, the same total number of deposition sites in either case is uniformly distributed on the whole surface of the bifurcation model.

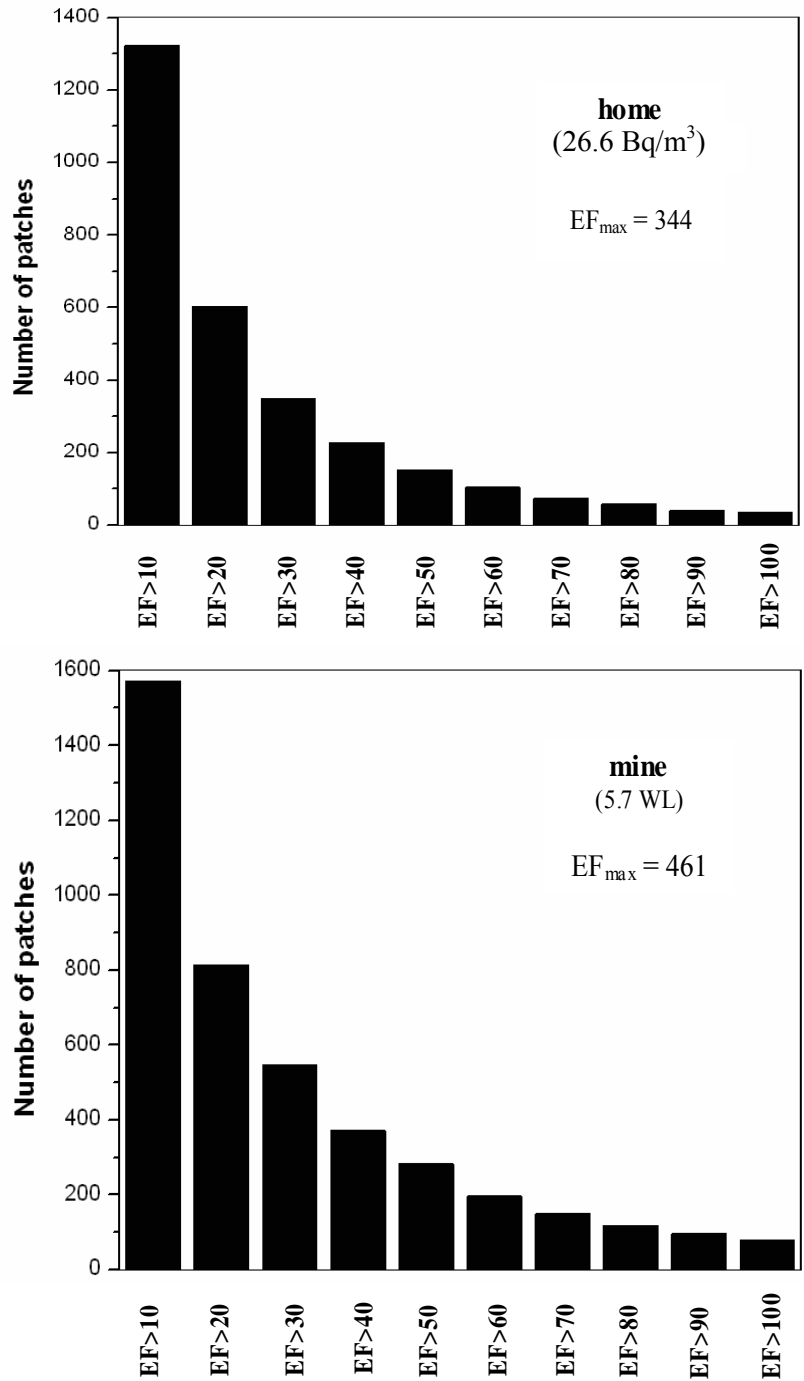

**Figure 2.** Distribution of deposition enhancement factors in homes (upper panel) and mines (lower panel). The simulation results are based on ten million inhaled particles in both environments. EFmax: maximum value of the enhancement factors, characteristic of the patch with the highest deposition density, WL: working level.

The figure demonstrates the highly inhomogeneous character of radon progeny deposition on bronchial airway surfaces. An important observation from a microdosimetric point of view is that in more than 1300 patches (1.73 % of the surface) for the home environment, the deposition density is at least one order of magnitude higher than its mean value. The total number of patches is about 75000. In 37 patches the deposition density of the radon progenies is even two orders of magnitude higher than the average deposition density. This suggests that for a few ten thousands of epithelial cells the radiation burden can be quite high even at low macroscopic or average doses. In uranium mines, this local burden can be even higher. In more than 1500 epithelial cell clusters (patches of the computational grid) the deposition density is more than ten times higher than the average density and 80 clusters are characterized by EFs exceeding 100. The number of particles deposited on the most exposed epithelial surface element is 461 times higher than the average number of the particles deposited on a patch. Primary hot spots have been found at the carinal ridge of the airway bifurcations, exactly at the locations where histological studies, e.g. Churg and Vedal (1996), have found high particle concentrations and neoplastic lesions. In following the term *hot spot* refers to areas where EF > 10. The distribution of hits and doses in the patch where EF has the maximum will be compared with the results of the uniform activity distribution models.

Figure 3 depicts the number of alpha hits for home and uranium mine environments as a function of the exposure level for uniform activity distributions and in the hot spots. In case of home (left panels) the average number of hits per bronchial cell nucleus is plotted versus the radon concentration at 70 years of exposure (lifetime) at uniform distribution and during a one-year exposure in the hot spot with maximum enhancement factor (EF = 344). The figure also demonstrates hit probability values presented in Table 2-1 of the BEIR VI Report with the two different averaging procedures based on volumetric ratios between basal and secretory cells presented in the Methods section. The volumetric differences between basal and secretory cells are presently not considered in the BEIR VI Report and the human respiratory tract model (HRTM) of the ICRP (1994). Assuming equal frequency and size for basal and secretory cell nuclei, the average hit probability is higher than if applying realistic weights, because there are about five times more basal than secretory cells. Thus the hit probability of the deeper lying basal cells is nearly four times higher than that of the secretory cells. The most likely reason why the BEIR VI Report apparently overestimates the hit probabilities by a factor of 2 (weighted average) or 3.3 (arithmetic average), is that the ICRP (1994) HRTM is a simple compartment model, which cannot consider the asymmetric distributions of the geometric parameters of the lung.

The average number of hits per bronchial cell nucleus in mines is presented as a function of WLM in Figure 3 again for a uniform distribution (right upper panel) and in the hot spots (right lower panel). The hit probabilities for a uniform deposition are compared with the corresponding values of BEIR VI Report. The figure reveals that in case of the most exposed cell surroundings, the number of hits per cell nucleus is about two orders of magnitude higher than at uniform distribution. Thus, the most exposed cell nuclei surroundings (EF = 344) receive about the same number of hits in one year as they would receive in a lifetime (70 years) assuming uniform deposition.

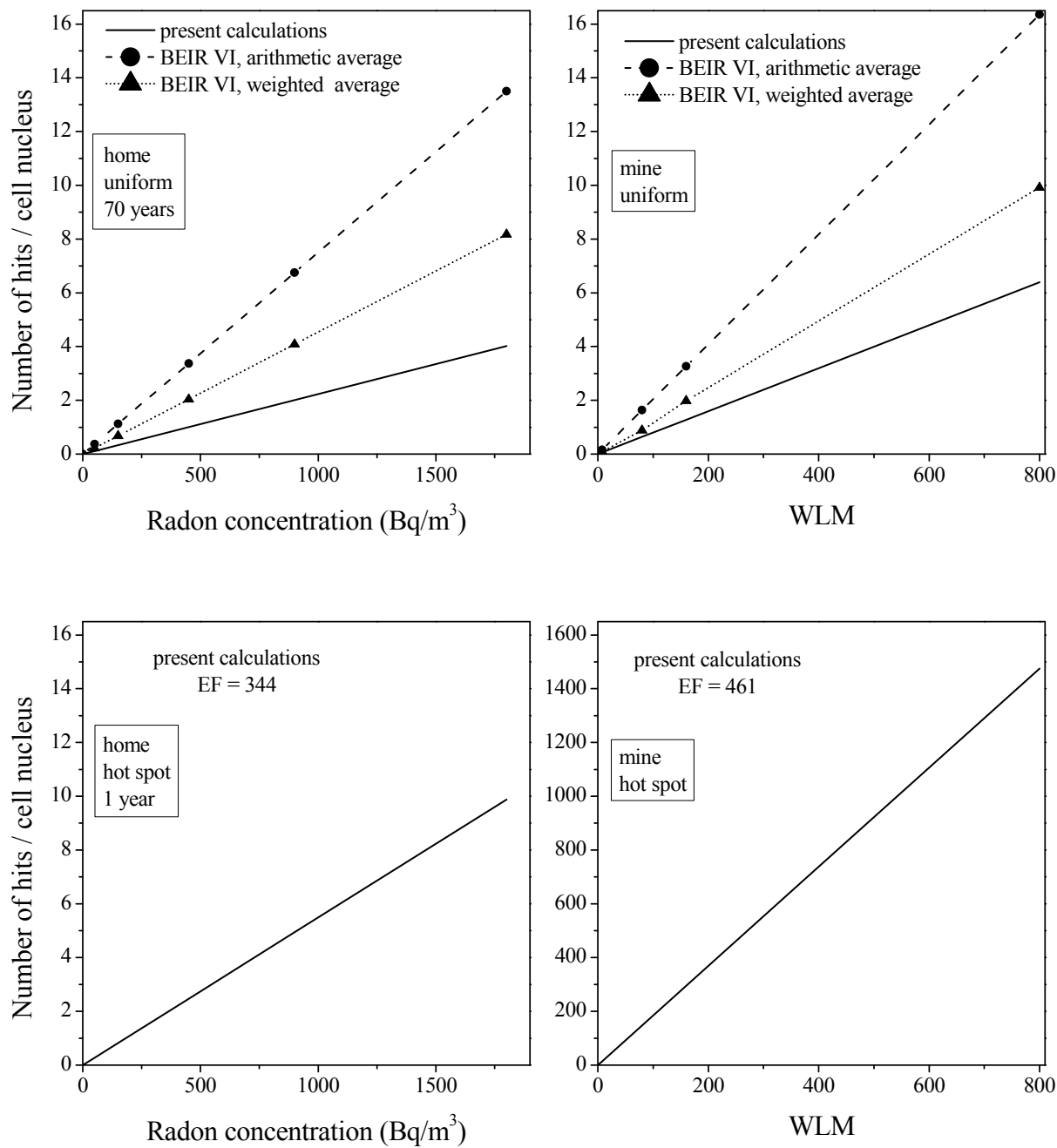

**Figure 3.** Average number of alpha particle hits per bronchial cell nucleus as a function of radon concentration in Bq/m$^3$ (for home exposure conditions over a period of 70 years) and the cumulative exposure in WLM (for mine exposure conditions), assuming uniform (upper panels) and hot spot distribution (bottom panels). Present hit calculations are compared to BEIR VI (1999) results, based either on the arithmetic average of hits to basal and secretory cells (●) or the average hits weighted by the relative fraction of basal and secretory cells (▲) in bronchial epithelium.

To study the effect of low doses, the inhaled radon concentrations during a certain period of time and WLM values were converted into bronchial cell nuclei doses using the conversion formulas presented in the Methods section.

Figure 4 illustrates the relationship between cell nucleus dose in hot spots and for a radon concentration during one year in homes (left panel) and WLM cumulative exposure in mines (right panel). The figure also demonstrates the nucleus doses in hot spots as a function of average cell nucleus dose. Based on BEIR VI Report (Table 3-5), assuming an equilibrium factor of 0.4 and an occupancy factor of 0.7 for homes, 250 Bq/m$^3$ burden during one year corresponds to 1 WLM indoor radon exposure. Assuming an equilibrium factor of 0.3 measured in the New Mexico uranium mine and 170 h/month mine occupancy, the corresponding conversion factor for mines is 1028 Bq/m$^3$/WLM. Assuming uniform deposition, the average dose received by a cell nucleus at 20 WLM is 96 mGy.

At 20 WLM radiation exposure, where the average cell nucleus dose is 96 mGy, the nucleus dose in the hot spots is extremely high: 16.5 Gy in homes and 22.1 Gy in mines [note: 20 WLM represents the lifetime residential indoor exposure (NRC 1991) or 18% of the average cumulative exposure received by a worker in New Mexico uranium mine (BEIR VI 1999)]. This means that in the hot spots the radiation burden can be very high even at low levels of exposure.

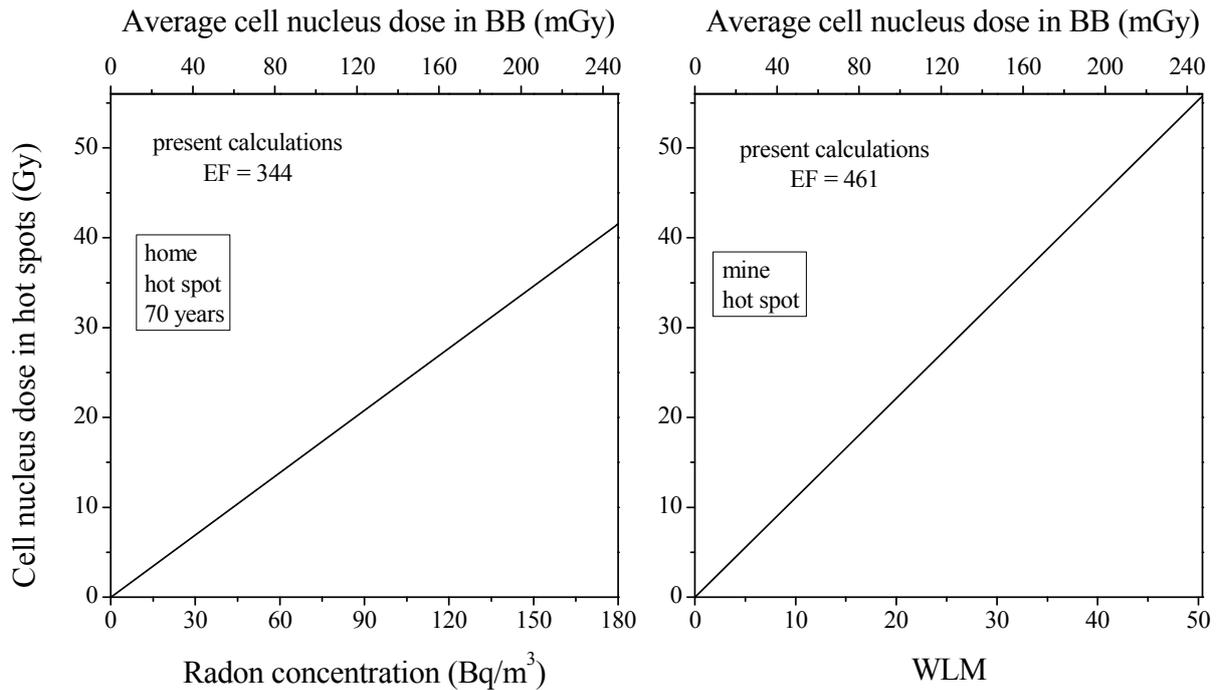

**Figure 4.** Illustration of the relationship between cell nucleus dose in hot spots and cumulative exposure, radon concentration and average cell nucleus dose, respectively, using exposure-specific maximum deposition enhancement factors.

Figure 5 illustrates the probability that a bronchial cell nucleus does not receive any hit, receives single hits, double hits, etc. for different exposures in homes (left panels) and mines (right panels). The hit probabilities are plotted assuming uniform surface activity distributions (upper panels) and in the hot spots of realistic inhomogeneous distributions (lower panels).

The frequency of multiple hits may be a key parameter regarding the late health effects, where the single cell traversals may be correlated to each other and hence may induce more complex cellular damages. In home environments, assuming uniform deposition and a 250 Bq/m$^3$ activity concentration, the probability of multiple hit over a 70 years period is less than 10 % and most probably the individual hits are well separated in time. In the hot spot, the probability of multiple hits at the same exposure level is about 30 % and practically 100 % for all radon concentrations greater than 1500 Bq/m$^3$ during one year. Assuming a uniform activity distribution and a mean cumulative exposure value of 110 WLM in the New Mexico uranium mine, the probability of multiple hits is about 25%. However in the hot spot exposure, this 25 % multiple hit probability is reached already at 0.5 WLM and the multiple hit probability is nearly 100 % for cumulative exposures greater than 4 WLM. In the New Mexico uranium mine, the probability of multiple hit is practically zero if assuming uniform activity distributions and 100 % in the hot spots for a working period of 1 month (5 WLM) which corresponds to the epithelial basal cell cycle time (BEIR VI 1999, Adamson 1985). It has been proposed by Leenhouts and Chadwick (1994) that in a protracted exposure the crucial parameter is the number of hits per cell cycle time and not over the whole exposure period. Thus, in the New Mexico uranium mine, practically all cells in hot spots receive multiple hits during a single cycle time, which may have serious biological consequences.

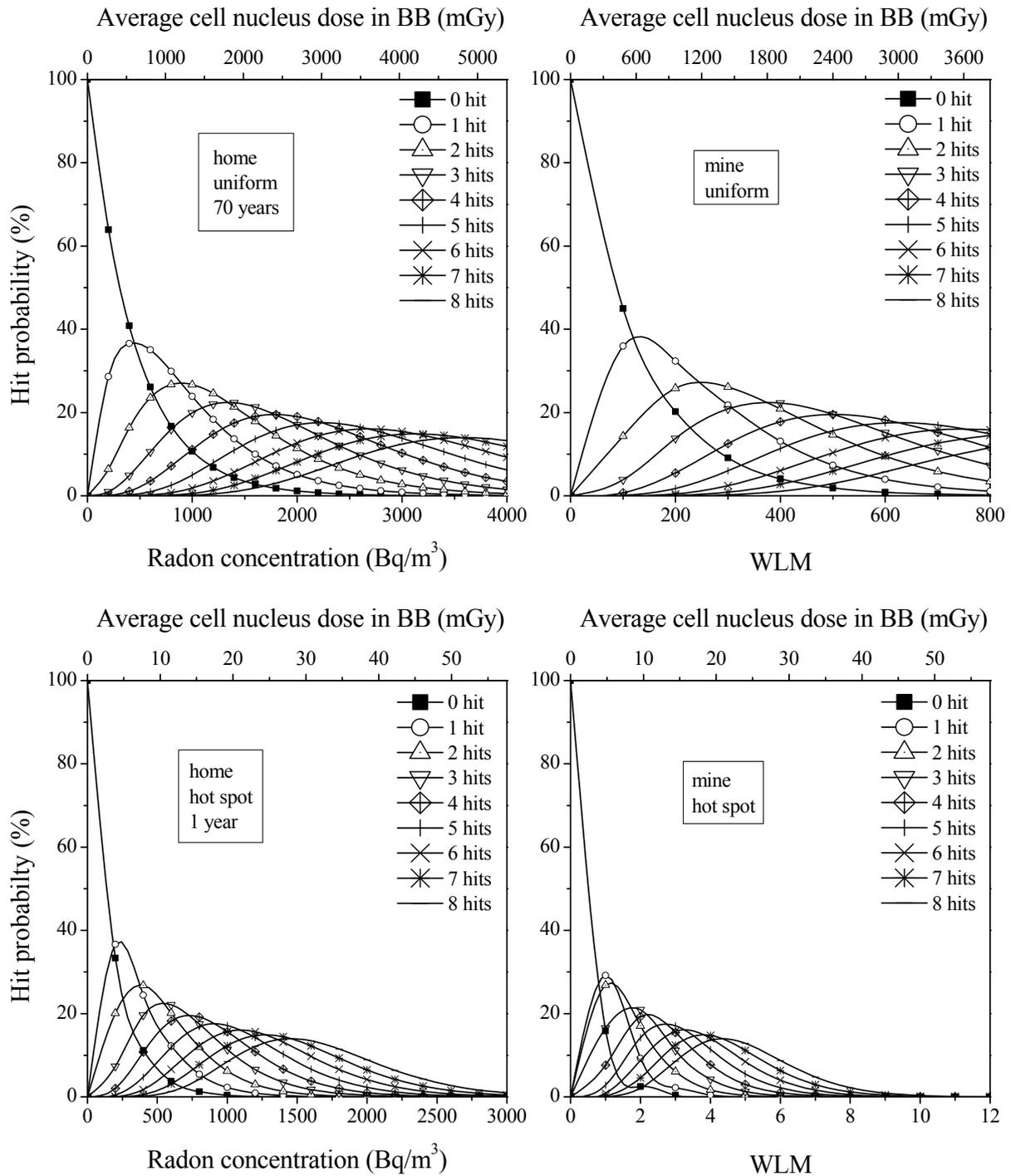

**Figure 5.** Alpha-hit probabilities as a function of radon concentration (home, left panels) and WLM (mine, right panels) assuming uniform activity distribution (upper panels) and in the hot spots (lower panels).

Alpha hit probabilities with the cell nuclei of the bronchial basal and secretory cells in the central airways in case of a uniform activity distribution (upper panels) and in hot spots (lower panels) are plotted in Figure 6 as functions of the radon concentration during one year in homes (left panels) and as a function of the cumulative exposure in WLM in mines (right panels), respectively. The figure also demonstrates how these hit probabilities are correlated with the average cell nucleus doses.

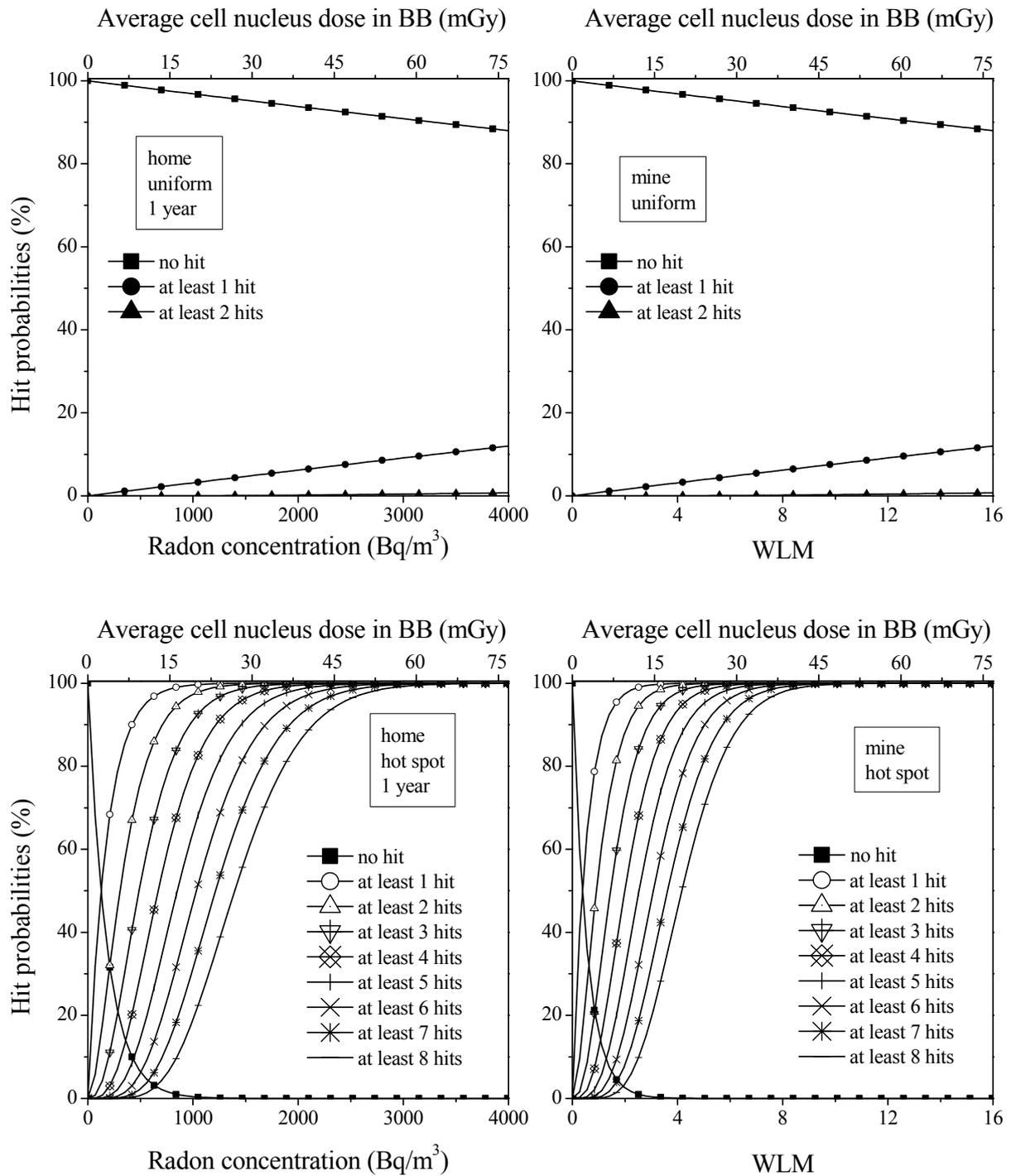

**Figure 6.** Hit probabilities for exposures in homes (left panels) and mines (right panels) for both uniform activity distributions (upper panels) and hot spots (lower panels).

The shapes of the curves are quite different at uniform distribution and in the hot spots of realistic distributions. Assuming a uniform distribution of activity (upper panels), which is presumed by current dosimetric models, there are practically no multiple hits and hit probabilities as a function of dose curve exhibit a linear shape in the low dose range. The results are quite the opposite in the case of hot spots of realistic deposition distributions, where practically all cells receive multiple hits and the hit probabilities as a function of dose curve are non-linear in the dose range of 0 - 75 mGy.

As it was mentioned, BEIR VI (1999) overestimates cellular hit probabilities by a factor of 2 or 3, depending on the averaging procedure, supposing uniform distribution of the deposited radon progenies (see Figure 3). BEIR VI is based on the ICRP compartment model and so cannot consider the asymmetric features of the lung. Since it does not consider the strong local inhomogeneity of deposition, the hit probabilities in the deposition hot spots are strongly underestimated by BEIR VI (1999) by a factor of 10 - 50.

Average hit numbers were also used for the estimation of cell transformation frequencies. Transformation probabilities under broad beam irradiation conditions were derived from *in vitro* experiments with C3H 10T1/2 cells (Miller et al. 1995, 1999). It is worth mentioning that in the hot spots an enhancement of apoptosis/necrosis might mitigate the effect of oncogenic mutations. From this point of view the experimental model of immortalized rodent embryo C3H 10T1/2 cells might not be fully representative of healthy human bronchial cells especially because they are presumably less prone to cell death than healthy cells. However, due to lack of pertinent information on *in vivo* response of human bronchial epithelial cells Miller's data were used. Transformation frequencies per surviving and per exposed cells are plotted in Figure 7 as functions of the average number of nuclear hits and related to indoor (left panel) and uranium mine exposure levels (right panel). The average numbers of hits for a uniform activity distribution and in hot spots of realistic distributions at different exposures were plotted in Figure 3. Applying it to the Miller et al. (1995, 1999) data, the cell transformation frequencies can be approximated by Figure 7. For example, the transformation frequency per surviving cells is about $6 \times 10^{-4}$, produced by the passage of six alpha particle through the cell nucleus, which corresponds to an indoor exposure of 50 Bq/m$^3$ over a period of 21.8 years in case of hot spot exposure, while transformation frequencies barely exceed background levels in case of a uniform activity distribution with the same radon concentration over a period of 52.1 years (see left panel). For uranium mine exposures, similar relationships are displayed in the right panel, indicating again that uniform activities may produce linear dose-effect relationships, while hot spot activities at the same exposure level will possibly lead to non-uniform dose-effect curves.

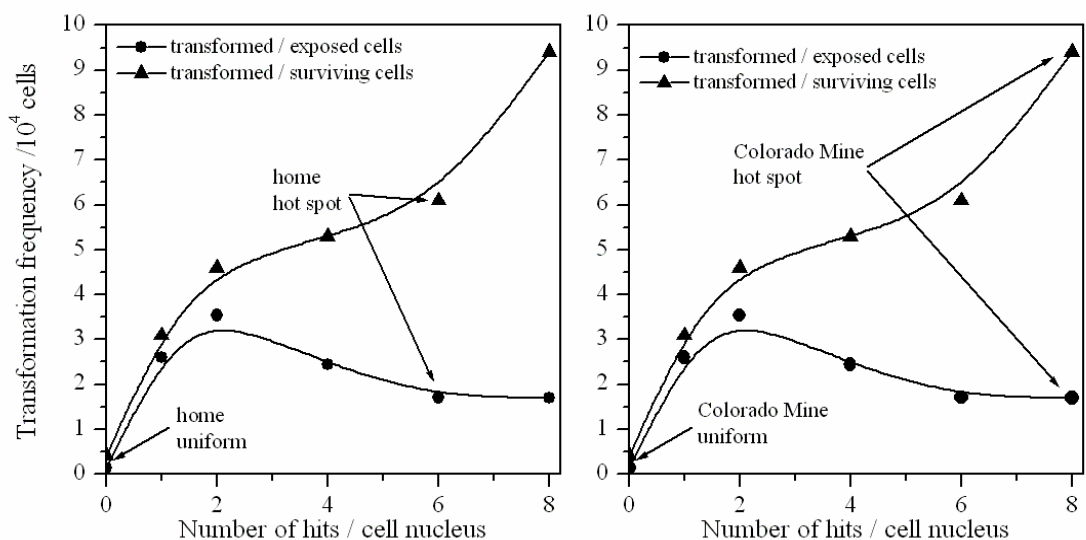

**Figure 7.** Transformation frequencies per surviving and per exposed cell for C3H 10T1/2 cells for a defined average number of alpha particles passing through the nucleus of these cells (Miller et al. 1999). In vitro transformation frequencies are related to in vivo home (left panel) and mine exposures (right panel) for both uniform and hot spot activities. Parameters: home, hot spot: 50 Bq/m$^3$, 21.8 years, EF = 344, cell nucleus dose = 3.6 Gy; home, uniform: 50 Bq/m$^3$, 52.1 years, cell nucleus dose = 50 mGy; mine, hot spot: 72426 Bq/m$^3$, 4.1 days, EF = 461, cell nucleus dose = 4.8 Gy; home, uniform: 72426 Bq/m$^3$, 9.8 days, cell nucleus dose = 50 mGy.

Alpha hit, and possibly subsequent transformation probabilities, of sensitive epithelium cells in the central airways increase linearly if presuming a uniform deposition of activity, but highly non-linearly if based on the computed, strongly inhomogeneous deposition of activity. The degree of inhomogeneity is so high that even in the low dose range thousands of sensitive epithelial cells receive high doses, which means that in case of radon inhalation the experienced health effects are related not to low but high dose biological mechanisms. ICRP Publication 103 (ICRP 103, 2007) defines the deterministic effects as clinically observable lesions after a sufficiently high number of cells died or turned into malfunction. It also states that deterministic effects always appear over a threshold dose. The high burden of some preferential areas suggest that such effects (e.g. inflammation) can occur at local scales even if the absorbed dose computed for the whole body or the whole lung is bellow say 100 mGy.

## 4. Conclusions

In the deposition hot spots of large central airways, where most of the radiation induced neoplastic lesions have been found, the local deposition density values can be a few hundred times higher than the average deposition density value. Here, the size of the unit surface element corresponds to the surface area of a few thousands of cells. The alpha particle hit probabilities in sensitive cell nuclei of the deposition hot spots are somewhat less but still more than hundred times higher than supposing uniform activity distributions of the inhaled attached and unattached radon progenies. The cellular hit probabilities can be quite high even at low doses in the vicinity of the carinal ridges of the central airways. The probabilities that an epithelial cell nucleus in the hot spots receives one or two or "n" hits increase exponentially and not linearly in the low dose range. Transformation frequences in the highly exposed areas are also orders of magnitude higher than the average transformation frequency.

**Acknowledgement**
This research was supported by the K61193 OTKA Hungarian Project and the EUREKA OMFB-445/2007 Project.